\documentclass{appolb}
\usepackage{epsfig}

\usepackage{subfigure}

\begin{document}
\title{Search for $\eta$-mesic nuclei at J-PARC%
\thanks{Presented at International Symposium on Mesic Nuclei, Cracow, 16 June 2010.}%
}
\author{Hiroyuki Fujioka
\address{Department of Physics, Kyoto University,\\
Kitashirakawa-Oiwakecho, Sakyo-ku, Kyoto 606-8502, Japan}
}
\maketitle
\begin{abstract}
The merits of searching for $\eta$-mesic nuclei by use of the ($\pi$, $N$) reaction at J-PARC is discussed, by comparison with an old experiment with the same reaction at BNL in 1980's. The importance of a pilot experiment using deuterium target is also stressed for a better understanding of the background which should exist even after the coincidence of the decay particles of $\eta$-mesic nuclei.
\end{abstract}
 
\section{Introduction}
It has been considered that a meson-nucleus system may exist as a quasi-bound state, mediated by strong interaction. For example, an antikaon attracts a nucleon very strongly in the isospin-0 channel, which is characterized by the existence of the $\Lambda(1405)$ resonance below the $\overline{K}N$ threshold. The possibility of a $\overline{K}$ nuclear bound state is suggested~\cite{Wycech86, Akaishi02}. Its decay width can be very narrow when the states lies below the $\Sigma \pi$ decay threshold, corresponding to $\sim 100\,\mathrm{MeV}$ binding of an antikaon. Experimental searches and reanalyses of past experiments have been performed, some of which reported observations of the simplest kaon-nuclear bound state, $K^-pp$~\cite{FINUDA05, DISTO10}, while the existence of such a state with its decay width narrow enough to be observed experimentally is still controversial. Further experiments will be planned at J-PARC, FOPI and DA$\Phi$NE, aiming to settle this problem.

As well as a kaonic nucleus, an $\eta$-mesic nucleus is also an interesting object to be investigated both theoretically and experimentally. The interaction between an $\eta$ meson and a nucleon is attractive, and the corresponding scattering length is estimated, with a large ambiguity, to be $(0.270\,\mbox{---}\,1.050) + (0.190\,\mbox{---}\,0.399)i\,\mathrm{fm}$\cite{Haider02}. While the first calculation by Haider and Liu~\cite{Haider86} indicated the existence of $\eta$-mesic nuclei for $A\ge 12$ by using the scattering length of $0.28+0.29i\,\mathrm{fm}$ or $0.27+0.22i\,\mathrm{fm}$, a larger scattering length may allow an $\eta$ meson to be bound in a lighter nucleus.

The reason for the large uncertainty of the scattering length arises from the difficulty to extract its information from experimental data; different from the case for charged pions or kaons, neither a scattering experiment in a direct way nor an X-ray measurement of mesic atoms is possible. Then, experimental observables such as the binding energy and the decay width of an $\eta$-mesic nucleus --- if it does exist and can be observed experimentally --- will provide us important information on $\eta N$ interaction.

It should be noted that the $N^*(1535)$ resonance plays an important role in near-threshold $\eta N$ interaction, since the resonance lies only $\sim 50\,\mathrm{MeV}$ above the $\eta N$ threshold, and it has a strong coupling with the $\eta N$ channel. If the properties of the resonance change at finite density, the $\eta N$ interaction will be different from that in free space, and this will affect the $\eta$-nucleus optical potential~\cite{Jido02}. Moreover, the main decay mode of an $\eta$-mesic nucleus will proceed via $\eta N\to N^*\to \pi N$.

Therefore, systematic investigations by using the same reaction on different nuclear targets to produce the known resonance of $N^*(1535)$ and $\eta$-mesic nuclei will be valuable for a better understanding of $\eta N$ interaction both in free space and in nuclear medium. This strategy is almost the same as a series of J-PARC experiments, to investigate the $\Lambda(1405)$ resonance by the ($K^-$, $n$) reaction on liquid deuterium target (E31 experiment~\cite{E31Prop}), and to search for a kaon-bound state $K^-pp$ by the same reaction on liquid helium-3 target (E15 experiment~\cite{E15Prop}).

In this paper, a possibility to use the ($\pi$, $N$) reaction, which may be feasible at J-PARC, will be discussed. 

\section{($\pi$, $N$) reaction at J-PARC}
The J-PARC accelerator complex, located in Tokai, Japan,  consists of a proton linac, and a $3\,\mathrm{GeV}$ rapid cycle synchrotron, and a $50\,\mathrm{GeV}$ main ring synchrotron. Intense proton beam is extracted from the main ring, and is transported to a production target, where secondary particles, such as pions and kaons, are produced. These particles are guided to experimental areas through beamlines.

As of summer 2010, there are two beamlines for charged particles, K1.8 and K1.8BR (branch line of K1.8), and the K1.1BR beamline is under construction. Their maximum momenta available are $\sim 2\,\mathrm{GeV}/c$, $1.1\,\mathrm{GeV}/c$, and $0.8\,\mathrm{GeV}/c$, respectively. Secondary beams, especially kaons, were successfully observed in both the K1.8 and K1.8BR beamlines in 2009. Already approved experiments, mainly on strangeness nuclear physics, are planned to start data taking at these beamlines. While pion beam with a proper momentum to investigate $\eta$-mesic nuclei can be extracted at either beamline, experimental plans by using the K1.8BR beamline will be discussed in the following.

\begin{figure}[t]
\begin{center}
\psfig{file=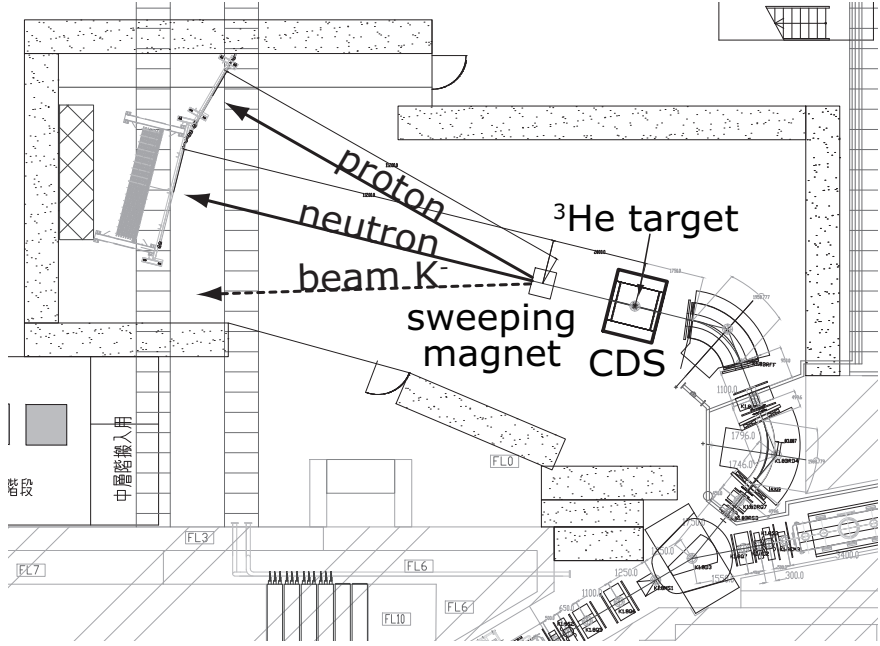,width=4.5in}
\caption{Setup for the J-PARC E15 experiment.}
\label{fig1}
\end{center}
\end{figure}

At the K1.8BR beamline, the preparation of the E15 experiment~\cite{E15Prop} is going on. It aims to search for a deeply-bound kaonic nuclear states $K^-pp$ by the $^3\mathrm{He}$ ($K^-$, $n$) reaction. Figure~\ref{fig1} shows the setup in the K1.8BR area. The K1.8BR beamline is branched from the K1.8 beamline at a dipole magnet, drawn in the lower right of the figure. The helium-3 target will be located near the end of the beamline, and will be surrounded by the Cylindrical Detector System (CDS). The CDS serves to detect the decay particles of the $K^-pp$ bound state into a $\Lambda$ and a proton. The CDS consists of a solenoid magnet with a magnetic field of $0.5\,\mathrm{T}$ parallel to the beam axis, the Cylindrical Drift Chamber (CDC), and the Cylindrical Detector Hodoscopes (CDH). The momentum of a charged particle is obtained by reconstructing its trajectory by the CDC. The CDH, which locates outside of the CDC, is used for the trigger purpose and the particle identification. Forward-scattered neutrons by the ($K^-$, $n$) reaction will be detected by an array of plastic scintillators, which will be put $15\,\mathrm{m}$ downstream from the target.

As well as the ($K^-$, $n$) reaction, the ($K^-$, $p$) reaction will be measured by adding plastic scintillators (P28 experiment~\cite{P28Prop}, approved as a part of the E15 experiment). Since a beam sweeping magnet will be located downstream of the target in the original E15 setup, scattered protons will be bent from the beam axis. Installation of drift chambers between the target and the magnet will help to reconstruct the reaction vertex, together with the trajectory of the scattered proton.

While this setup is optimized for the E15/P28 experiments, a measurement of the ($\pi$, $N$) reaction is also possible with a small modification. As discussed below, the detection of the decay particles of $\eta$-mesic nuclei and $N^*(1535)$ is very important for improving the signal-to-noise ratio. The usage of the CDS around the target is very suitable for this purpose.
 
\section{Search for $\eta$-mesic nuclei}
The first experiment to search for $\eta$-mesic nuclei was carried out at BNL by using the ($\pi^+$, $p$) reaction on lithium, carbon, oxygen, and aluminum targets~\cite{Chrien88}. The spectrometer was tilted by $15^\circ$ against the beam axis, which was the best condition according to a theoretical suggestion by Liu and Haider~\cite{Liu86}. 

However, a peak structure corresponding to $\eta$-mesic nuclei was not observed for any of the targets.
It does not necessarily mean that the ($\pi$, $N$) reaction is not a proper way to produce $\eta$-mesic nuclei.

\begin{figure}[t]
\begin{center}
\psfig{file=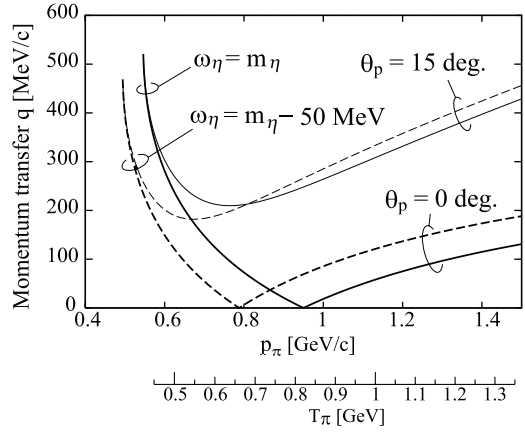,width=2.75in}
\caption{Momentum transfer for the $A(\pi,\ N)$ reaction (heavy mass limit) as a function of incident pion momentum ($p_\pi$) or kinetic energy ($T_\pi$), for different scattering angles ($\theta_p$) of $0^\circ$ and $15^\circ$. The solid (dashed) line corresponds to the $\eta$ energy at $\omega_\eta=m_\eta$ ($m_\eta-50\,\mathrm{MeV}$). This figure is taken from Ref.~\cite{Nagahiro09}.}
\label{fig4}
\end{center}
\end{figure}

First, the momentum transfer is as large as $200\,\mathrm{MeV}/c$ for the scattering angle of $15^\circ$ (see Fig.~\ref{fig4}). Nagahiro \textit{et al.} calculated the missing-mass spectra for the same reaction with different scattering angle ($0^\circ$ and $15^\circ$)~\cite{Nagahiro09}. For the former case, the magic momentum exists, so that the momentum transfer can be zero. Then, only substitutional states are excited, because the angular momentum of the target nucleus should be conserved. On the other hand, many subcomponents of different configurations of a nucleon-hole and an $\eta$ meson will be populated, and they will hinder the observation of a possible contribution of $\eta$-mesic nuclei.

Secondly, the inclusive spectrum Chrien \textit{et al.} obtained would contain a huge background from multi-pion production \textit{etc.}, irrelevant to $\eta$ production. If they had used a decay counter for coincidence analysis, they might get a spectrum with a better signal-to-noise ratio.

For these reasons, there is still room to consider an experiment with the same reaction, which satisfies the recoilless condition. An exclusive measurement, \ie the detection of decay particles of $\eta$-mesic nuclei ($\eta N\to N^*(1535) \to \pi N$), as the TAPS/MAMI experiment~\cite{Pfeiffer04} and the COSY-GEM experiment~\cite{COSYGEM09} have adopted, is mandatory.

We proposed to study the ($\pi^-$, $n$) reaction\footnote{This is essentially the same as the ($\pi^+$, $p$) reaction because of the isospin symmetry. In the LoI, we limited the discussion to the ($\pi^-$, $n$) reaction only, taking into account merits from an experimental point of view, but the other reaction may be feasible as well.} on $^7\mathrm{Li}$ and $^{12}\mathrm{C}$ targets at J-PARC~\cite{Itahashi_LoI}, motivated by recent theoretical calculations on this reaction~\cite{Jido02, Nagahiro09}. They focused on in-medium property of $N^*(1535)$, to which an $\eta$ meson and a nucleon in an $\eta$-mesic nucleus couples; the $\eta$-nucleus optical potential is attractive (replusive) when $\omega +m_N^*(\rho) - m_{N^*}^*(\rho)$  is negative (positive), where $\omega$ is the $\eta$ energy, $m_N^*(\rho)$ and $m_{N^*}^*(\rho)$ are effective masses of $N$ and $N^*$ at nuclear density $\rho$, respectively. In free space, since the mass gap of $N^*$ and $N$, i.e. $m_{N^*}-m_{N}$, is larger than the $\eta$ meson mass, $\eta$-nucleus interaction is attractive at threshold. If the mass gap reduces at finite density and becomes smaller than the $\eta$ meson mass, the interaction near threshold will be repulsive. In the chiral doublet model, which treats a negative-parity $N^*(1535)$ baryon as the chiral partner of the nucleon, the mass gap is expected to reduce as a function of nuclear density $\rho$. To the contrary, the mass gap will not change largely at finite density according to the chiral unitary model, since $N^*(1535)$ is regarded as a dynamically generated resonance of meson-baryon scattering, with a large contribution of the $K\Sigma$ channel, and a $\Sigma$ baryon is free from the Pauli blocking in a nucleus.

The beam momentum will be around $0.7\mbox{--}1.0\,\mathrm{GeV}/c$ so as to satisfy the recoilless condition. The scattered neutron (proton) will be detected by zero-degree detectors like the E15/P28 experiment. The decay particles of $\eta$-mesic nuclei, such as a proton and a $\pi^-$, will be also identified by a decay counter.

However, even though a proton and a $\pi^-$ are detected in coincidence, there still remain background events which do not associate with an $\eta$ meson. In order to estimate the background level in the actual experiment, a pilot experiment to investigate the elementary process by use of deuterium target, $\pi^++d\to p+p^*(1535)$, because the $N^*(1535)$ serves as a doorway in forming $\eta$-mesic nuclei. The background mentioned above may correspond to an event which does not produce a $p^*(1535)$ baryon in the $\pi^++d$ reaction. The detail of the $\pi^++d$ reaction will be discussed in the next section.

\section{pilot experiment: $d$($\pi^+$, $p$) reaction}
\begin{figure}[t]
\begin{center}
\subfigure[double-scattering reaction]{\psfig{file=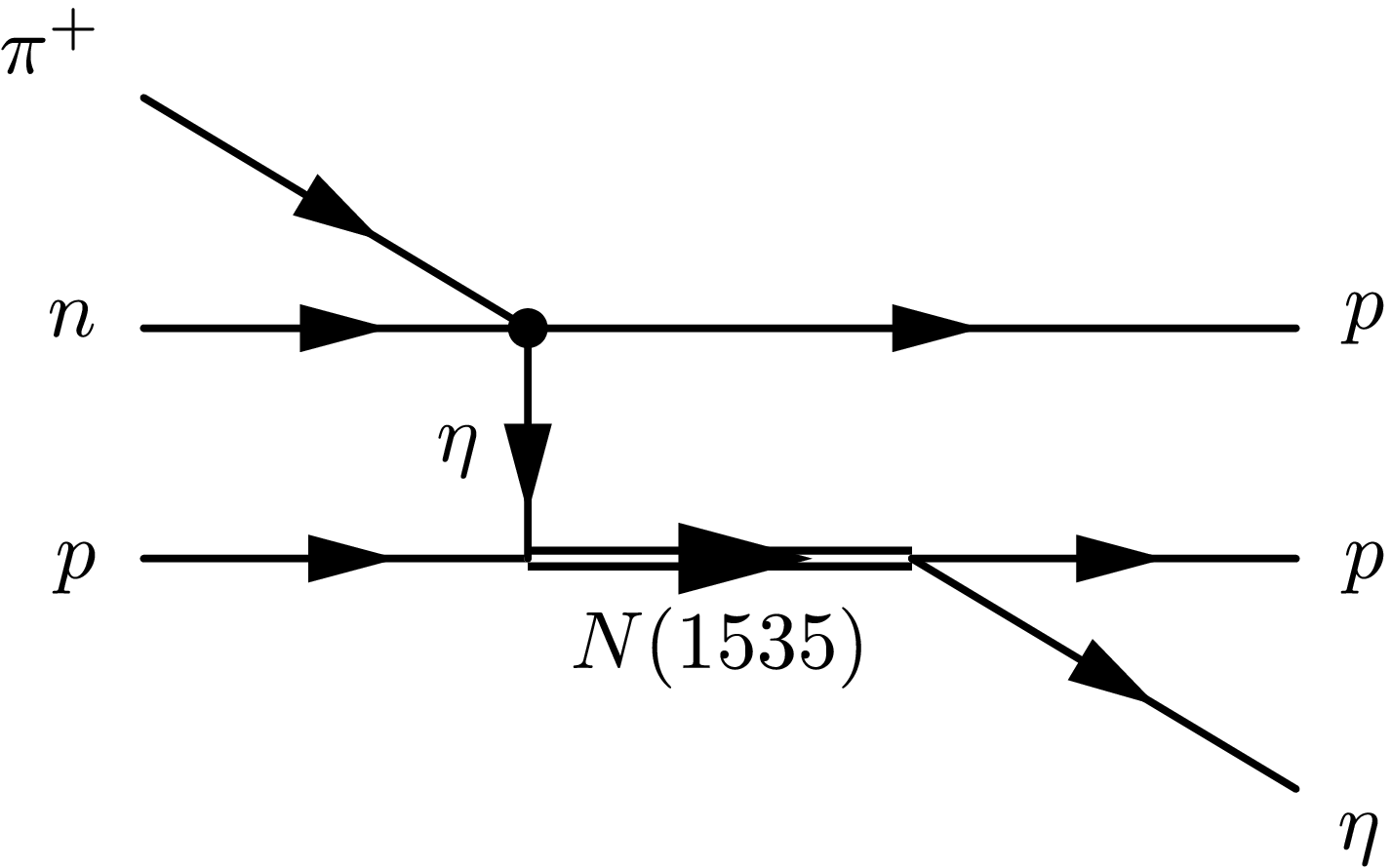,scale=0.35}\label{fig2a}}
\hspace{0.5cm}
\subfigure[quasi-free reaction]{\raisebox{9.9mm}{\psfig{file=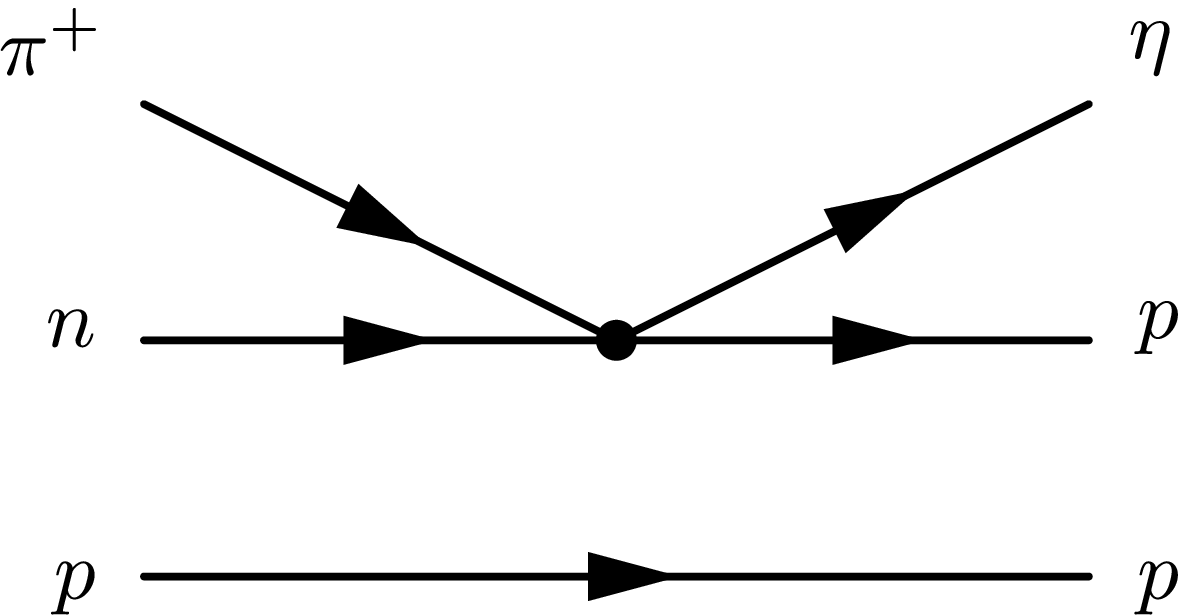,scale=0.35}}\label{fig2b}}
\caption{Diagrams for the $\pi^+d\to pp\eta$ reaction.}
\end{center}
\end{figure}

A clear signal to identify the production of the $N^*(1535)$ baryon is to identify the $N\eta$ decay, since its branching ratio is 45--60\%~\cite{PDG08}, significantly larger than that from other $N^*$ baryons. Therefore, the cross section of the reaction:
\begin{equation}
\pi^++d \to p+p+\eta
\label{re1}
\end{equation}
will give a good estimation of the $N^*(1535)$ production cross section. It should be noted that this final state can be originated not only from a double-scattering reaction (Fig.~\ref{fig2a}), but also from a quasi-free reaction (Fig.~\ref{fig2b}), and their contributions should be distinguished experimentally. The double-scattering reaction is a two-step process, the elementary process of $\pi^+ +n\to p+\eta$, followed by the rescattering of the $\eta$ meson, and it is similar to the formation of $\eta$-mesic nuclei by the ($\pi$, $N$) reaction, except for the target of the rescattering process with the $\eta$ meson. As the spectator proton in the quasi-free reaction has a small momentum, the detection of two protons in the final state\footnote{Protons slower than $\sim 200\,\mathrm{MeV}/c$ in the laboratory system easily stops in the target itself, which means most of spectators protons can not be detected without some effort.} will suppress the contribution of the quasi-free process.

\begin{figure}[t]
\begin{center}
\psfig{file=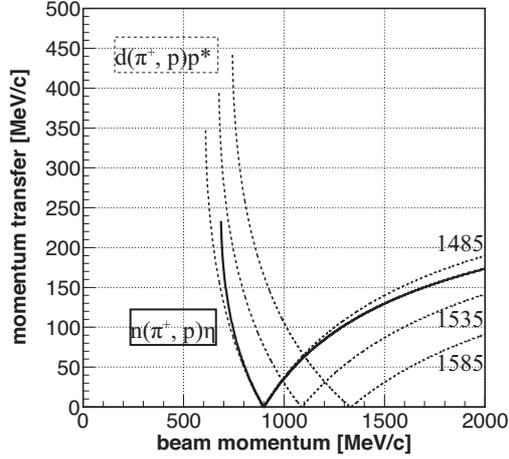,width=2.6in}
\caption{Momentum transfer for the $n$($\pi^+$, $p$)$\eta$ reaction (solid line) and the $d$($\pi^+$, $p$)$p^*$ reaction (dashed lines) for three different $p^*$ masses (1485, 1535, 1585$\,\mathrm{MeV}/c^2$).}
\label{fig3}
\end{center}
\end{figure}

Figure~\ref{fig3} shows the momentum transfer for this reaction with the scattering angle $0^\circ$, together with that for the quasi-free reaction. When the beam momentum is around $0.8\mbox{--} 1.1\,\mathrm{GeV}/c$, the momentum of the $N^*(1535)$ baryon is small.

On the other hand, the reaction with a different final state:
\begin{equation}
\pi^++d \to p+p+\pi^0
\label{re2}
\end{equation}
corresponds to the actual measurement about $\eta$-mesic nuclei. Some of the $N^*(1535)$ baryon via the double-scattering reaction can decay into $p\pi^0$, but other reactions, such as the rescattering of $\pi$ or $\rho$ mesons, will contribute to this final state. 

If we compare the differential cross section of the reactions (\ref{re1}) and (\ref{re2}) for the same forward-going proton momentum, or the same energy of $p\eta$ or $p\pi^0$ system, we will be able to estimate the contribution of $N^*(1535)$ in the reaction (\ref{re2}), and at least a qualitative evaluation of the background level in the $A$($\pi$, $N$) reaction to produce $\eta$-mesic nuclei will be possible.

Furthermore, the cross section of the double-scattering reaction is sensitive to low-energy $\eta N$ interaction (see Fig.~\ref{fig2a}), because a quasi-elastic scattering of an $\eta$ meson and a proton takes place. Therefore, the differential cross section of the reaction (\ref{re1}) may provide independent information on this interaction, which has been investigated by photoproduction and $\pi$-induced reactions.

\begin{figure}[t]
\begin{center}
\psfig{file=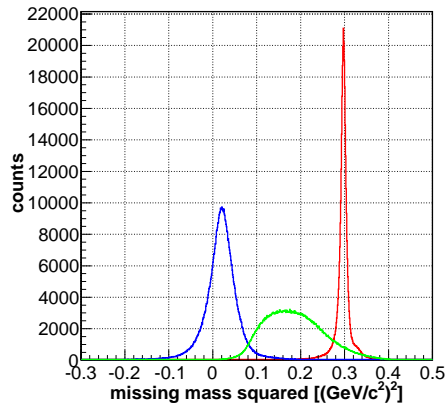,width=2.6in}
\caption{Missing mass squared for the $d$($\pi^+$, $pp$)$X$ reaction for $X=\pi^0$, $(\pi\pi)^0$, and $\eta$ (from left to right). The peak position is in good agreement with the PDG mass squared of the corresponding particle.}
\label{fig5}
\end{center}
\end{figure}

Th experimental setup for this pilot experiment is still under consideration. If we can use the E15/P28 setup, the detection of only two protons is enough to distinguish the possible final states of $pp\eta$, $pp\pi^0$, and $pp(\pi\pi)^0$, by the missing-mass analysis for $d$($\pi^+$, $pp$)$X$. Figure~\ref{fig5} is a preliminary result of a Monte Carlo simulation, taking into account the detector resolution These contributions are well-separated from each other.

\section{Conclusion}
The ($\pi$, $N$) reaction, once investigated by Chrien \textit{et al.} at BNL, is revisited. An exclusive measurement with the recoilless condition may enhance a signal from $\eta$-mesic nuclei. Before starting such an experiment, a pilot experiment with the $d$($\pi^+$, $p$) is also planned. It aims to estimate the contributions of the $N^*(1535)$ production and irrelevant backgrounds by detecting the $pp \eta$ and $pp \pi^0$ states.

\section*{Acknowledgments}
The author would like to appreciate Dr. K.~Itahashi, Dr. H.~Nagahiro, Dr. D.~Jido, and Prof. S.~Hirenzaki for valuable discussions.

\end{document}